\documentclass[a4paper]{article}
\usepackage[top=5cm,bottom=2.5cm,headheight=90pt,headsep=15pt,left=5.5cm,right=1.5cm,marginparwidth=4.2cm,marginparsep=0.5cm]{geometry}

\usepackage{marginnote}
\reversemarginpar  % sets margin notes to the left
\usepackage{lipsum} % Required to insert dummy text
\usepackage{calc}
\usepackage{siunitx}
\usepackage{lineno}
\usepackage[utf8]{inputenc}
\usepackage[T1]{fontenc}
\usepackage[
backend=biber,
natbib=true,
sortcites=true,
defernumbers=true,
style=authoryear,
citestyle=authoryear-comp,
maxnames=99,
maxcitenames=2,
giveninits=true,
terseinits=true,
url=false
]{biblatex}
\DeclareNameAlias{default}{family-given}
\renewbibmacro{in:}{%
  \ifentrytype{article}{}{\printtext{\bibstring{in}\intitlepunct}}} % remove 'In:' before journal name
\DeclareFieldFormat[article]{pages}{#1} % remove pp.
\AtEveryBibitem{\ifentrytype{article}{\clearfield{number}}{}} % don't print issue numbers
\DeclareFieldFormat[article, inbook, incollection, inproceedings, misc, thesis, unpublished]{title}{#1} % title without quotes
\usepackage{csquotes}
\RequirePackage[english]{babel} % must be called after biblatex
\DeclareBibliographyCategory{ignore}
\addtocategory{ignore}{recommendation} % adding recommendation to 'ignore' category so that it does not appear in the References
\usepackage{nameref}
\usepackage[pdfborder={0 0 0}]{hyperref}  % sets link border to white
\usepackage{graphbox}  % loads graphicx ppackage with extended options for vertical alignment of figures
\usepackage{microtype}
\DisableLigatures[f]{encoding = *, family = * }
\RequirePackage[default,scale=0.90]{opensans}
\usepackage{xcolor}
\definecolor{darkgray}{HTML}{808080}
\definecolor{mediumgray}{HTML}{6D6E70}
\definecolor{ligthgray}{HTML}{d9d9d9}
\definecolor{pciblue}{HTML}{74adca}
\definecolor{opengreen}{HTML}{77933c}
\usepackage{changepage}
\usepackage[aboveskip=1pt,labelfont=bf,labelsep=period,singlelinecheck=off]{caption}

\usepackage{fancyhdr}  % custom headers/footers
\usepackage{lastpage}  % number of page in the document
\fancyhfoffset[L]{4.5cm}  % offsets header and footer to the left to include margin
\renewcommand{\headrulewidth}{\ifnum\thepage=1 0.5pt \else 0pt \fi} % header ruler only on first page

\newcommand{\DOIrecommendationlink}{\href{https://doi.org/\DOIrecommendation}{https://doi.org/\DOIrecommendation}}

\newcommand{\PCI}{Peer Community In Neuroscience}

\newcommand{\beginingpreprint}{
\vspace*{0.5cm}
\begin{flushleft}
\baselineskip=30pt
\marginpar{
\large\textnormal{\color{pciblue}\\RESEARCH ARTICLE}\\
\vspace*{0.5pt}
\\
\includegraphics[align=c,width=0.5cm]{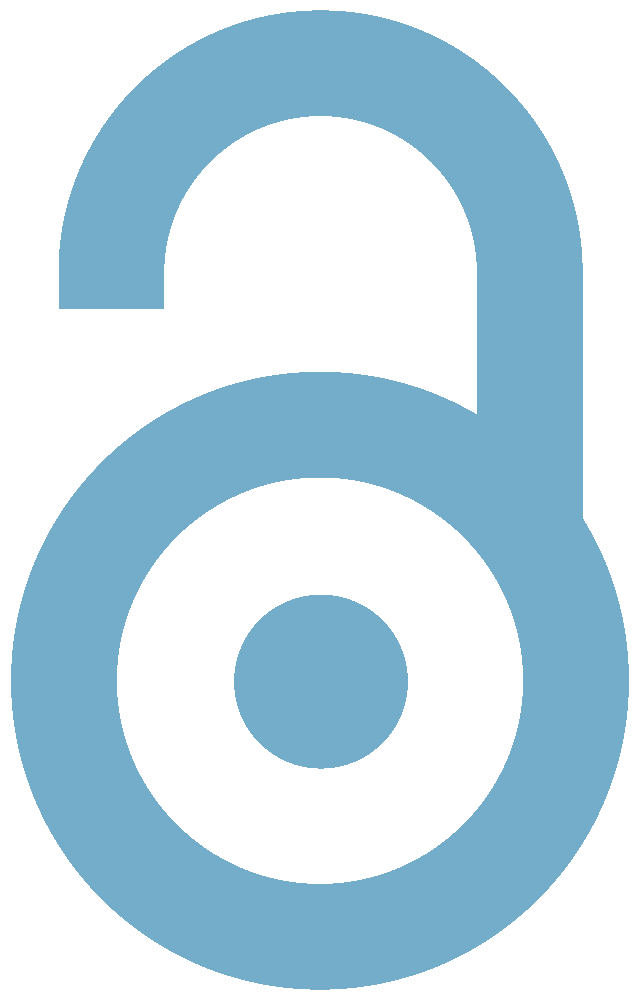} \space \large\textbf{\color{pciblue}Open Access}\\
\\
\includegraphics[align=c,width=0.5cm]{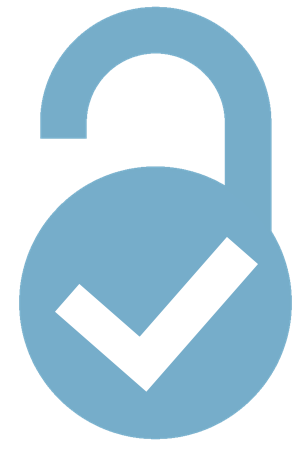} \space \large\textbf{\color{pciblue}Open Peer-Review}\\
\\
\includegraphics[align=c,width=0.5cm]{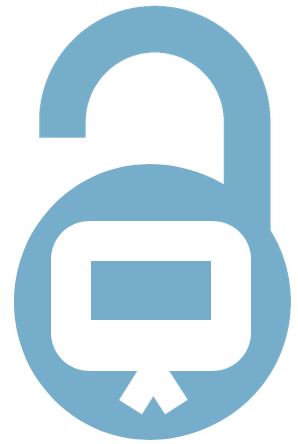} \space \large\textbf{\color{pciblue}Open Code}\\
\\
\\
\\
\\
\raggedright
\scriptsize\textbf{Cite as:}\space
\citeas\\
\vspace*{0.5cm}
\textbf{Posted:} \datepub\\
\vspace*{0.5cm}
\textbf{Recommender:}\\
\recommender\\
\vspace*{0.5cm}
\textbf{Reviewers:}\\
\reviewers\\
\vspace*{0.5cm}
\textbf{Correspondence:}\\
\href{mailto:\email}{\email}\\

}
{\huge
\fontseries{sb}\selectfont{\preprinttitle}}
\end{flushleft}
\vspace*{0.25cm}
\begin{flushleft}
\Large
\listauthors
\end{flushleft}
\bigskip
{\raggedright
\listinstitutions}
\begin{flushleft}
\fcolorbox{lightgray}{lightgray}{
\parbox{\textwidth - 2\fboxsep}{
\centering\large{\fontseries{sb}\selectfont{This article has been peer-reviewed and recommended by 
\emph{\PCI} (\DOIrecommendationlink)}}\\
}}
\end{flushleft}
\vspace*{0.5cm}
\fcolorbox{pciblue}{pciblue}{
\parbox{\textwidth - 2\fboxsep}{
\vspace{0.25cm}
\textbf{\large{\textsc{Abstract}}}\\
\preprintabstract\\

\footnotesize{\textbf{\emph{Keywords: }}\preprintkeywords}
\vspace{0.25cm}}
}
}

\usepackage{booktabs}       % professional-quality tables
\usepackage{amsfonts}       % blackboard math symbols
\usepackage{nicefrac}       % compact symbols for 1/2, etc.
\usepackage{microtype}      % microtypography
\usepackage{amsmath}
\usepackage{amssymb}
\usepackage{graphicx}
\usepackage{float}
\usepackage{mathtools}
\usepackage{svg}
\usepackage{bm}
\usepackage{chngcntr}

\newcommand{\RR}{\mathbb{R}}
\newcommand{\B}[1]{\mathbf{#1}} % math bold face
\newcommand*{\tran}{^{\mkern-1.5mu\mathsf{T}}} % pretty transpose
\newcommand{\thinkron}{\! \otimes \!} % 'thin' kronecker operator with less white-space
\newcommand{\HB}[1]{\hat{\mathbf{#1}}} % hat bold face
\newcommand{\DB}[1]{\dot{\mathbf{#1}}} % dot bold face

\newcommand{\preprinttitle}{Nonlinear computations in spiking neural networks through multiplicative synapses}
\newcommand{\listauthors}{\raggedright 
Michele Nardin\textsuperscript{1}, \space
James W Phillips, \
William F Podlaski\textsuperscript{2} \&
Sander W Keemink \textsuperscript{2,3}
}
\newcommand{\listinstitutions}{
\textsuperscript{1} Institute of Science and Technology Austria, Klosterneuburg, Austria 
\\
\textsuperscript{2} Champalimaud Research, Champalimaud Centre for the Unknown, Lisbon, Portugal
\\
\textsuperscript{3} Artificial Intelligence, Donders Institute for Brain, Cognition and Behaviour, Radboud University, Nijmegen, the Netherlands
}
\newcommand{\datepub}{November 18th, 2021}
\newcommand{\recommender}{Marco  Leite}
\newcommand{\DOIrecommendation}{10.24072/pci.cneuro.100003}
\newcommand{\reviewers}{Two anonymous reviewers}
\newcommand{\citeas}{Michele Nardin, James W Phillips, William F Podlaski, Sander W Keemink. (2021) Nonlinear computations in spiking neural networks through multiplicative synapses. ArXiv, ver. 4 peer-reviewed and recommended by Peer Community in Neuroscience. https://arxiv.org/abs/2009.03857v4}
\newcommand{\email}{michele.nardin@ist.ac.at; sander.keemink@donders.ru.nl}
\newcommand{\preprintabstract}{The brain efficiently performs nonlinear computations through its intricate networks of spiking neurons, but how this is done remains elusive. While nonlinear computations can be implemented successfully in spiking neural networks, this requires supervised training and the resulting connectivity can be hard to interpret. In contrast, the required connectivity for any computation in the form of a linear dynamical system can be directly derived and understood with the spike coding network (SCN) framework. These networks also have biologically realistic activity patterns and are highly robust to cell death. Here we extend the SCN framework to directly implement any polynomial dynamical system, without the need for training. This results in networks requiring a mix of synapse types (fast, slow, and multiplicative), which we term multiplicative spike coding networks (mSCNs). Using mSCNs, we demonstrate how to directly derive the required connectivity for several nonlinear dynamical systems. We also show how to carry out higher-order polynomials with coupled networks that use only pair-wise multiplicative synapses, and provide expected numbers of connections for each synapse type. Overall, our work demonstrates a novel method for implementing nonlinear computations in spiking neural networks, while keeping the attractive features of standard SCNs (robustness, realistic activity patterns, and interpretable connectivity). Finally, we discuss the biological plausibility of our approach, and how the high accuracy and robustness of the approach may be of interest for neuromorphic computing.}
\newcommand{\preprintkeywords}{Spiking Neural Networks; Multiplicative Synapses; Nonlinear Dynamical Systems; Direct Derivation}
\begin{document}
\beginingpreprint

%%%%%%%%%%%%%%%%%%%%%%%%%%%% Text of the preprint %%%%%%%%%%%%%%%%%%%%%%%%%%%%%%
%%%%%%%%%%%%%%%%%%%%%%%%%%%%%%%%%%%%%%%%%%%%%%%%%%%%%%%%%%%%%%%%%%%%%%%%%%%%%%%%
% use \\ and a blank line to mark the end of paragraph and starting of a new one

\section{Introduction}
A central quest in neuroscience is to understand how the brain's neural networks are able to perform the computations needed to solve complex tasks. One promising hypothesis is that networks represent relatively low-dimensional signals (compared to network size) \parencite{cunningham2014dimensionality, keemink2019decoding}, and in this lower-dimensional space implement nonlinear dynamical systems through recurrent connectivity \parencite{eliasmith_unified_2005, mante2013context, sussillo2014neural, abbott2016building, barak2017recurrent, mastrogiuseppe2018linking}. The resulting networks usually achieve nonlinear computation through a basis-function approach: non-linearities at various levels (neural \parencite{jaeger2001echo,maass2002real,eliasmith_unified_2005, mastrogiuseppe2018linking}, synaptic \parencite{thalmeier2016learning}, or dendritic \parencite{thalmeier2016learning, alemi2018learning, abbott2016building}) are weighted to achieve a given computation through supervised training. The task of achieving nonlinear computation is then off-loaded to the basis-functions and the training method, and as a result the link between network connectivity and computation may be unclear. Additionally (unlike real biological systems), the resulting models do not always exhibit robustness to perturbations \parencite{li_robust_2016}, and suffer from unrealistic activity levels~(e.g. \parencite{eliasmith_large-scale_2012}) compared to expected levels~\parencite{barth_experimental_2012}. In contrast, the recurrent connectivity required for any \textit{linear} dynamical system can be directly defined  (i.e., without any training) for a spiking neural network through the theory of spike coding networks (SCNs) \parencite{boerlin2013predictive}. SCNs are consistent with many features from biology, such as sparse and irregular activity, robustness to perturbations (such as cell death) \parencite{barrett2016optimal, calaim2020robust}, and excitation/inhibition-balance \parencite{boerlin2013predictive, deneve2016efficient}. Could we use a similar analytical approach to introduce nonlinear computations?
\\

In SCN theory, fast recurrent connections are used to efficiently and accurately maintain a stable internal representation. Any linear dynamical system can then be directly implemented through the addition of slower recurrent connections \parencite{boerlin2013predictive}, which will drift the internal representation according to the desired dynamics. While nonlinear dynamics have previously been implemented in SCNs, this was achieved through the aforementioned basis-function approach \parencite{abbott2016building, thalmeier2016learning, alemi2018learning}. Here we extend the original SCN derivation for linear dynamics \parencite{boerlin2013predictive}, by directly deriving the connectivity required for any polynomial dynamical system. The resulting networks require an additional set of slow connections with multiplicatively interacting synapses, and we thus term our model \emph{multiplicative} SCNs, or mSCNs.
\\ 

We demonstrate the capability of mSCNs through a precise implementation of the Lorenz system, as well as an implementation of a double pendulum. While polynomial systems can in principle approximate any other system \parencite{de1959stone}, this can quickly become infeasible, as higher-order polynomial systems require higher-order synapses (pair-wise, triplets, quadruplets, etc.), resulting in a dense and complicated all-to-all connectivity structure. We address the need for higher-order synapses by demonstrating how higher-order computations can be approximated by successive network layers with solely pair-wise synapses. Additionally, we demonstrate that the assumption of all-to-all connectivity can be loosened if each neuron is selective only for a subset of the relevant variables for the computation.
\\

Our theory of mSCNs harnesses all the appealing properties of previous SCN implementations (in particular robustness to cell-death and irregular spiking activity), but now includes a directly derivable and more interpretable connectivity structure for a large class of nonlinear dynamical systems. Finally, the efficiency and accuracy of our networks might be of use for neuromorphic applications, especially for representing dynamical systems that are well-described by lower-order polynomials.

\section{Spike coding networks}

Here and in the following sections we present the main results and refer the reader to the Methods section for details. Consider a network of $N$ spiking neurons, which emit spike trains of the form $s_i(t) = \sum_k \delta(t^i_k - t)$, where $\delta$ is the Dirac delta function and $\{t^i_k \geq 0 \}$ is the set of discrete times at which a spike was emitted. The population spike train is described by the vector $\B s(t) = [s_1(t), \dots, s_N(t)]\tran$. Vectors will be denoted by lower case bold letters, and wherever possible we will omit the time index for the sake of text clarity.

\subsection{Linear autoencoder}
Suppose that a given $K-$dimensional signal $\B x(t) \in \RR^K$ should be represented by the output activity of the network (Fig.~\ref{Fig1}A). How should the neurons then spike to accomplish this task? The theory of spike coding networks approaches this through two core assumptions \parencite{boerlin2013predictive,deneve2016efficient,barrett2016optimal}: \\

\textit{(a) Linear decoding:} 
the network representation $\hat{\B x}(t)$ is read out as
\[
\hat{\B x} = \B D\B r % \approx \B x
\]
where $\B D \in \RR^{K \times N}$ is the decoding matrix, and $\B r(t) =  [r_1(t), \dots, r_N(t)]\tran$ are filtered spike-trains
\[ 
\dot{\B r} = -\lambda \B r + \B s,
\]
where $\lambda$ is the leak time-constant. The variable $\B r$ can be seen as a neuron's time-dependent rate, or equivalently the effect of a neuron's spikes on the post-synaptic potential of other neurons. \\

\textit{(b) Efficient spiking:}
$\B D_i$ (the $i-$th column vector of the matrix $\B D$) represents the contribution that a spike from neuron $i$ will have on each dimension of the read-out signal; more specifically, a spike at time $t$ will update the current readout as $\hat{\B x}(t) \to \hat{\B x}(t) + \B D_i$. Assumption (b) requires that this spike should only occur if it improves the read-out. Formally, we require that a spike reduces the $\ell^2$-error between the readout and the signal. Thus, neuron $i$ will fire at time $t$ if and only if
\[
\left\Vert\B x(t) - (\hat{\B x}(t) + \B D_i)\right\Vert_2^2 < \left\Vert\B x(t) - \hat{\B x}(t)\right\Vert_2^2 .
\] 
After some algebra (see Methods~\ref{Methods:GeneralDerivation}), and defining the membrane potential of each neuron as $V_i = \B D_i\tran(\hat{\B x} - \B x)$, one finds an underlying dynamical description of the system where each neuron spikes whenever $V_i > T_i$, with $T_i =  \B D_i\tran \B D_i / 2$, and membrane potential dynamics 
\begin{equation} \label{eq_repr}
\dot{\B v} = -\lambda \B v + \B D\tran(\dot{\B x} + \lambda \B x) - \B D\tran \B D \B s,
\end{equation}
where $\B v = (V_1, \dots, V_N)$. Thus, starting from the two core assumptions, we have derived a recurrently connected network of leaky integrate-and-fire neurons. Through recurrent connections (given by $\B D\tran \B D$) the network accurately tracks its input signal $\B x$,  (Fig.~\ref{Fig1}B+C top row).

\subsection{Linear dynamics}
\begin{figure}
\centering
\includegraphics[width=\textwidth]{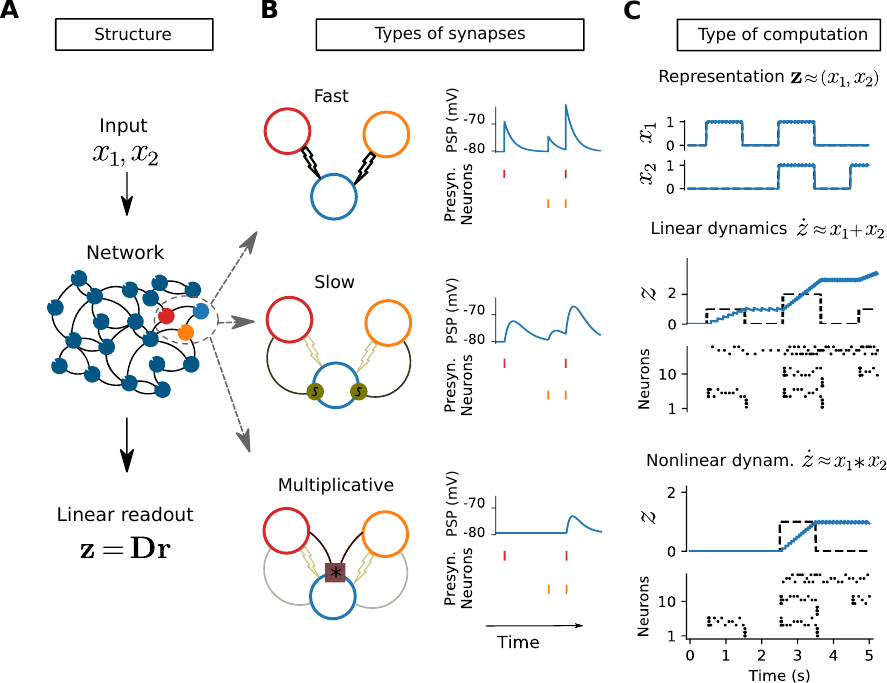}
\caption{Multiplicative Spike Coding Networks (mSCNs) can implement polynomial dynamics. 
\textbf{(A)} Schematic representation of the network.
\textbf{(B)} The network has three types of synapses, illustrated for two neurons (red and orange) connecting to another (blue). The postsynaptic potential (PSP) of a cell endowed with multiplicative synapses will be affected only if the two presynaptic neurons fire very close in time to each other. Higher order multiplicative interactions can also be necessary, and are illustrated in Supp.~Fig.~\ref{SupFigHighOrder}.
\textbf{(C)} Example computations enabled by the different types of synapses: (top) the network represents the two inputs ($x_1$ and $x_2$). The blue line represents the output of the network, the dashed black lines the real input. (Middle) A network which computes the dynamical system $\dot z = x_1 + x_2$. The black dashed line represents the real sum of the inputs, the blue line represents the output of the network. (Bottom) A network which computes the nonlinear dynamical system $\dot z = x_1*x_2$, and thus integrates the product of $x_1$ and $x_2$. For both linear and nonlinear dynamics the fast synapses are also required.
}
\label{Fig1}
\end{figure}

In the above derivation, the signal $\B x$ was provided directly to the network, but this is not strictly necessary. If $\B x$ follows some known linear dynamics $\dot{ \B x} = \B A \B x$ (with $A \in \RR^{K\times K}$) then its trajectory can be computed by the network \parencite{boerlin2013predictive}. Their derivation uses the fact that $\B x \approx \hat{\B x}$, so that (\ref{eq_repr}) can be approximated as
\begin{equation} \label{eq_lindyn}
\begin{split}
\dot{\B v} &= -\lambda \B v + \B D\tran\left(\B A \hat{\B x} + \lambda \hat{\B x}\right) - \B D\tran \B D \B s \\
      &= -\lambda \B v + \B \Omega_f \B s + \B \Omega_s \B r,
\end{split}
\end{equation}
where so-called ``fast'' connections $\B \Omega_f = - \B D\tran \B D$ keep the error constrained on a short time-scale (Fig.~\ref{Fig1}B+C top row), and ``slow'' connections $\B \Omega_s = \B D\tran (\B A + \lambda \B I)\B D$ implement the dynamical computation using the filtered spikes $\B r$ (Fig.~\ref{Fig1}B+C middle row). Here ``fast'' and ``slow'' refer to the rise-time of the synaptic PSPs (Fig.~\ref{Fig1}B). While technically an approximation, this implementation works well in practice and can closely reproduce a given linear dynamical system (Fig.~\ref{Fig1}B+C middle row).

%%%%%%%% insert figure 1 here %%%%%%%%%%%
%%%%%%%% edn figure 1 %%%%%%%%%%%%%%%%%

\section{Nonlinear dynamics}
The approximation in (\ref{eq_lindyn}) was originally conceived for linear systems, but can in principle be extended to any arbitrary dynamical system $\dot{\B x} = F(\B x)$ (with  $F: \RR^K \to \RR^K$). The full network dynamics then become
\begin{equation} \label{eq2} 
\dot{\B v} = -\lambda \B v + \B{D}\tran(F(\hat{\B x}) + \lambda \hat{\B x}) - \B{D}\tran \B{D} \B s,
\end{equation}
with the problem that the nonlinear function $F()$ has to be somehow computed by the network (or individual neurons). Previous work approximated this computation through a set of basis functions (\parencite{thalmeier2016learning, alemi2018learning}; see Methods~\ref{Methods:BasisFuns}), which can be interpreted as dendritic nonlinearities. Here we take a different approach. We note that any smooth nonlinear function can in principle be approximated by a polynomial transformation~\parencite{de1959stone}.
%\footnote{BP: could be an interesting paper to parencite?: \href{https://ieeexplore.ieee.org/abstract/document/80265}{https://ieeexplore.ieee.org/abstract/document/80265}.}
Furthermore, any polynomial function $F:R^K \to R^K$ containing terms with maximum degree $g$ can be written in the form
\begin{equation}\label{kronpoly}
    % F(\B x) = \sum_{d=0}^g \B A_d \thinkron^d (\B x)
    F(\B x) = \sum_{d=0}^g \B A_d \B x^{\otimes d},
\end{equation}
where $\B A_d \in \RR^{K \times K^d}$ is the matrix of coefficients for the polynomials of degree $d$, and we define 
%$\thinkron$ represents the Kronecker product
% $\otimes^d(\B M)$ to be a function that returns $\B M \thinkron \B M \thinkron \cdots \thinkron \B M$ applied $d$ times, with the convention that $1$ is returned for $d=0$ and $\B M$ for $d=1$. 
$\B M^{\otimes d} = \B M \thinkron \B M \thinkron \cdots \thinkron \B M$ as the Kronecker product applied $d$ times, with the convention that $\B M^{\otimes 0} =
1$ and $\B M^{\otimes 1} = \B M$. The Kronecker product is closely related to the outer-product and computes all possible pair-wise multiplications between the elements of two matrices. For example, the Kronecker product of two vectors of length $l$ is itself a new vector of length $l^2$ (see Methods~\ref{Methods:kronecker} for a detailed explanation).

Using this notation, the connectivity and dynamics for the \emph{multiplicative} SCN network implementing a polynomial function can be directly derived as

\begin{equation} \label{eq_nonlindyn} 
\begin{split}
\dot{\B v} &= -\lambda \B v - \B D\tran \B D \B s + \B D\tran(\sum_{d=0}^g \B A_d \B D^{\otimes d} \B r^{\otimes d} + \lambda \hat{\B x}) \\
      &= -\lambda \B v + \B \Omega_f \B s + \B \Omega^{m0}_s + \B \Omega^{m1}_s \B r + \B \Omega^{m2}_s \B r^{\otimes 2} + \dots + \B \Omega^{mg}_s \B r^{\otimes d} \\
      &= -\lambda \B v + \B \Omega_f \B s + \sum_{d=0}^g \B \Omega^{md}_s \B r^{\otimes d},
\end{split}
\end{equation}
where $\B \Omega_f = - \B D\tran \B D$, $\B \Omega_s^{m1} = \B D\tran (\B A_1 + \lambda \B I)\B D$ and $\B \Omega^{md}_{s} = \B D\tran \B A_d \B D^{\otimes d}$ for $d \in \{0,2,3,\dots,g\}$. \\

With this factorization we demonstrate how any given multiplicative interaction of state-variables can be accurately implemented through multiplicative synapses between neurons (Fig.~\ref{Fig1}B+C~bottom, Methods~\ref{Methods:LinearNonlinearSCNS}). The matrix $\B \Omega^{md}_s$ then  represents $d-$th degree multiplicative interactions between cells. In particular, $\B \Omega^{m2}_s$ represents the connectivity required for each cell to multiply each pair of their inputs (Fig.\ \ref{Fig1}B+C bottom row), with the synapse essentially acting as a coincidence detector. Higher order synapses would behave similarly but with three or more coincident spikes (Supp.~Fig.~\ref{SupFigHighOrder}).  While these higher order interactions are unlikely to be biologically feasible, lower order multiplicative interactions may indeed be possible in biology, and have been hypothesized before \parencite{koch_multiplying_1992}, as we will discuss further in the discussion. In principle, increasingly complex nonlinear dynamics may be implemented through the inclusion of higher-order terms in eq.~(\ref{eq_nonlindyn}) ($\B \Omega_{md}$ for $d>2$), though this flexibility comes with increased cost on the number of synapses and neural interactions. In a later section we will show how to avoid interactions beyond pair-wise synapses, and we will derive the expected number of connections for each type of synapse. \\

Compared to linear dynamics, multiplicative synapses enable nonlinear computations such as AND gates (Fig.~\ref{Fig1}C bottom). Overall, the above derivation demonstrates that the presence of multiplicative synapses arises naturally in mSCNs from extending the spike coding framework to polynomial dynamical systems.

\subsection{Lorenz attractor}

We illustrate the functionality of the mSCN formalism through an implementation of a simple dynamical system, the Lorenz attractor. The Lorenz attractor is a system of ordinary differential equations first studied by Edward Lorenz, which may lead to chaotic solutions \parencite{lorenz1963deterministic, strogatz2018nonlinear}. It is defined as
\begin{align*}\label{lorenz}
\dot x &= \sigma (y - x) \\
\dot y &= x (\rho - z) - y \\
\dot z &= x y - \beta z.
\end{align*}
Notably, this system contains pairwise multiplicative terms of the state-variables, thereby making it a polynomial (and nonlinear) dynamical system. In the following, we use the ``classical'' parameter values $\sigma =10$, $\beta =8/3$ and $\rho =28$, for which the system is in a chaotic regime. This is a useful case study as the resulting behavior is very sensitive to small representation errors, and has previously been used to test spiking network dynamical system implementations (e.g. \parencite{thalmeier2016learning}). \\

\begin{figure}
\centering
\includegraphics[width=\textwidth]{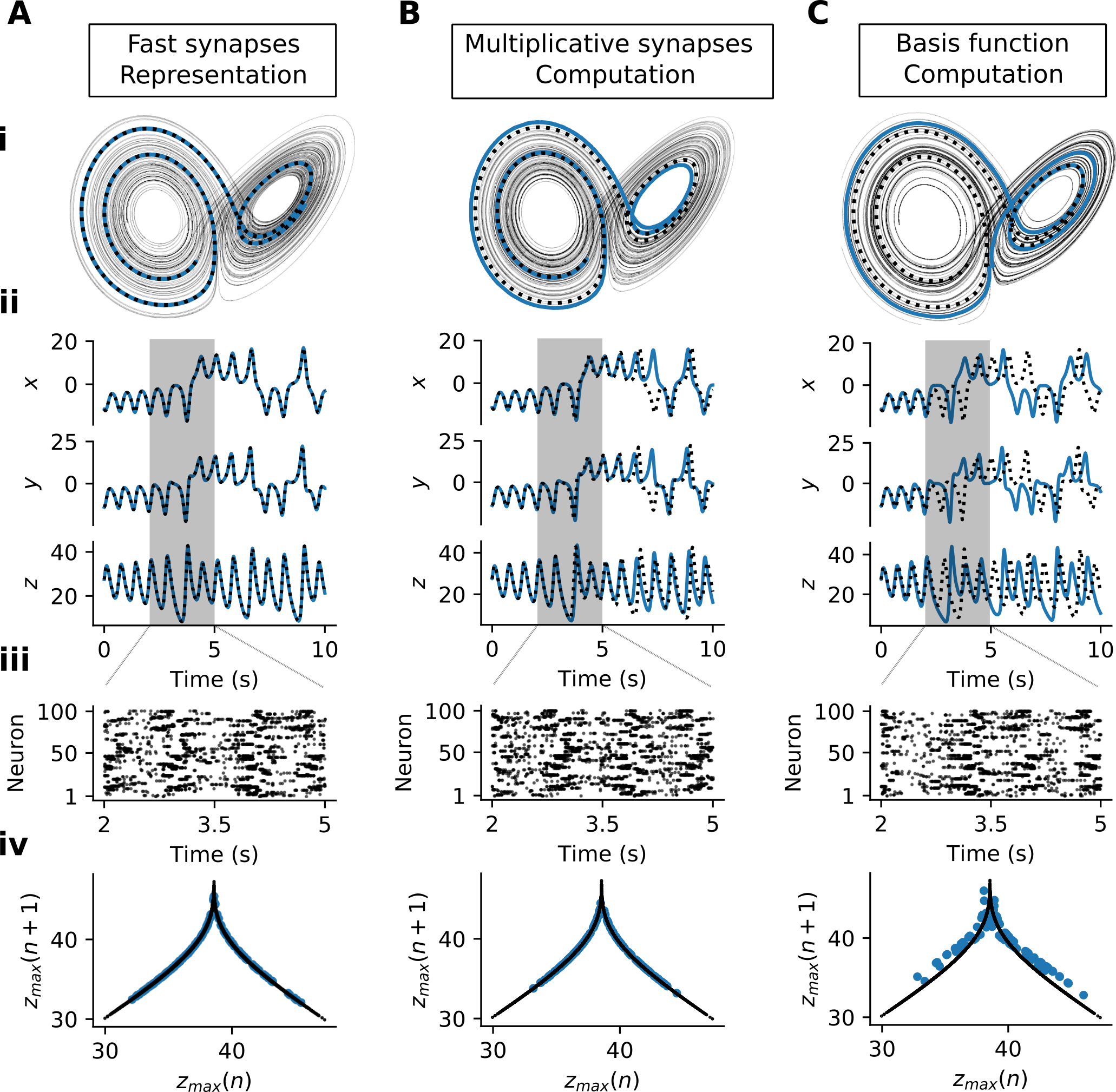}
\caption{Implementation of a Lorenz dynamical system. Across columns: 
\textbf{(A)} The numerical solution was found using an explicit Runge-Kutta method of order 4. The network contains only fast synapses and receives as input the output of the Lorenz simulation, which the network read-out tracks closely.
\textbf{(B)} A network with multiplicative synapses (mSCN) computing the Lorenz attractor through its network dynamics. 
\textbf{(C)} A network with nonlinear basis functions computing the Lorenz attractor through its network dynamics. Across rows:  
\textbf{(i)} 3D view of the network readout for 100 sec (grey). The dotted line shows the `true' simulation in the 2-5sec period, and the blue line shows the corresponding network output trajectory.
\textbf{(ii)} Each network readout dimension (blue) across time vs the `true' solution (black dotted). The gray region indicates the 2-5sec period used in panels i and iii.
\textbf{(iii)} Raster plot with spikes emitted by the neurons in the time interval 2 - 5 sec. 
\textbf{(iv)} Peak analysis over 100 sec: blue = network output, black = `true' Lorenz simulation.
}
\label{Fig2}
\end{figure}

We implemented the Lorenz system in three ways, each using networks of $N=100$ neurons. First, we simulated the Lorenz system in a standard numerical simulation (Runge-Kutta method), and fed the dynamic variables $\B x$ directly as input into an autoencoding network with only fast synapses. Note that the correct trajectory is thus continuously being fed into this network. This control acted as an upper-bound on the accuracy of representation with a spiking network of a given size and read-out weight magnitudes. As expected the network represented the system with high fidelity --- only small deviations arose compared to the standard numerical simulation due to the discrete spiking representation of the network (Fig.~\ref{Fig2}Aii). To have a better idea of the accuracy of the representation, we followed \parencite{thalmeier2016learning} and compared the values of neighboring peaks in the dynamics of the $z$ variable, which closely tracked a function defined by the pure Lorenz simulation (Fig.\ \ref{Fig2}Aiv). \\

Next, we implemented the Lorenz system in an mSCN (following Eq.~\ref{eq_nonlindyn}, see Methods~\ref{Methods:Lorenz}). The resulting network is able to compute the Lorenz dynamics with high accuracy (Fig.\ \ref{Fig2}B). The representation tracked the dynamics of the standard numerical simulation (dotted lines) for a reasonable amount of time, despite the attractor's chaotic nature (Fig.\ \ref{Fig2}B~i+ii), though this depends on the simulation time step (set to $0.1$ms here). Additionally, despite the deviations from the `true' trajectory, the peak analysis demonstrates that qualitatively the implementation is near perfect (Fig.~\ref{Fig2}B~iv). Furthermore, the network simulation still displays the extreme robustness to cell death of traditional SCNs (Supp. Fig.~\ref{SupFigRobust}). \\

Lastly, for comparison's sake, we implemented the same dynamical system using basis functions with trained weights (Methods~\ref{Methods:BasisFuns}) in order to understand the benefits and drawbacks of each approach. We used $500$ basis functions per neuron. The Lorenz attractor is again qualitatively well reproduced (Fig.~\ref{Fig2} top). But in contrast to the two previous schemes, the implementation with basis functions led to more inaccuracies (Fig.\ \ref{Fig2}C), quickly resulting in missed shifts in the dynamics. This was likely due to the approximate nature of the basis function implementation, and we note that more precise simulations might be possible with more and different basis functions (see e.g. Fig.~3f-h in \parencite{thalmeier2016learning}; though even there, outliers are still present). \\

These results might suggest that the direct implementation of the Lorenz system with the mSCN is capable of more accurate dynamics than a basis function implementation. However, we note three caveats here. First, the accuracy of the dynamics depends on the nature of the underlying system --- e.g., small inaccuracies would matter less for a system with stable fixed points. 
Second, the Lorenz attractor is perfectly described by a polynomial, and other dynamical systems might be better described by a basis-function implementation (of similar complexity as a given mSCN). 
Third, the differences in accuracy and scaling of the two implementations suggests that each may be more suitable depending upon the specific problem at hand (on the order of $N^3$ parameters are needed for the mSCN and $bN^2$ for the basis function implementation, where $b$ is the number of basis functions). 

%%%%%% Figure 2 (Lorenz) here %%%%%%%%%%%%%
%%%%%%%% end figure 2 %%%%%%%%%%%%%%%%%%%%%%

\section{Higher-order polynomials with sequential networks}

While the Lorenz system is a good case study for demonstrating the power and accuracy of mSCNs, a core problem remains: higher order polynomials necessitate higher order multiplicative interactions. E.g., a polynomial of order $3$ would require a $\B r^{\otimes 3}$ term, with on the order of $N^4$ synapses.
Such precise higher-order interactions may not always be feasible, either biologically or on a neuromorphic substrate. However, as we show here, this is not strictly necessary. Across populations, it is possible to combine many sequential pairwise interactions to achieve multiplications of any other order.

\subsection{Input transformations}
In principle, an SCN can also represent a nonlinear transformation of a signal $G(\B x)$, where $G$ is a smooth function $G: \RR^K \to \RR^M$, $M\geq 1$. For this, consider a new SCN with decoder $\B W$, spikes ${\bm\sigma}$, and filtered spike trains ${\bm\rho}$. 
In that case $\B v$ evolves according to 
\begin{equation}\label{Nonlin_repr_vec}
    \DB v = -\lambda \B v
            +  \B W\tran\left( \B{J}_G(\B x) \dot{\B x} + \lambda G(\B x)\right)
            - \B W\tran \B W \B {\bm\sigma},
\end{equation}
where $\B{J}_G$ is the Jacobian of $G$ (Methods~\ref{Methods:GeneralDerivation}). 
%But the transformed signal then needs to either be provided to the network, or computed in some way. 
The problem here is that the transformation function has to be either provided or computed by the network.
%As we showed for dynamical systems above, 
However, if $G$ is a polynomial function, the computation can be implemented using multiplicative synapses --- e.g., for quadratic terms, we can use $G(\B x) = \B x \thinkron \B x$.
Specifically, we need a network that takes as input the output of another network $\HB x = \B D \B r$ (with spikes $\B s$) and returns $\B W \B {\bm\rho} \approx \HB x ^ {\otimes 2}$ (Methods~\ref{Methods:RepresentingMultiplication}). Using the fact that $\dot{\B r} = -\lambda \B r + \B s$, we obtain dynamics
\[
\DB v = -\lambda \B v  + \B \Omega_{x} (\B r \otimes \B s + \B s \otimes \B r + - \lambda \B r \otimes \B r) + \B \Omega_f^W \B {\bm\sigma},
\]
with $\B \Omega_x = \B W\tran (\B D \otimes \B D)$ and $\B \Omega_f^W = - \B W \tran \B W$. We illustrate the resulting network and its inputs and outputs in Supp.~Fig.~\ref{SupFigKronProd}.

\subsection{Combining networks}
Now, the combination of a standard mSCN (eq. \ref{eq_nonlindyn}) with a network which calculates the square of its inputs (Supp.~Fig.~\ref{SupFigKronProd}), results in a system with the ability of computing third-order multiplications with only pairwise (second-order) synapses (Fig.~\ref{Fig4}A). Notably, one network computes the pairwise multiplications, and the other computes the desired third-order dynamic equation using another pairwise multiplication of the squared network output ($\B x \thinkron \B x$) and $\B x$. \\

More concretely, we consider a polynomial function $F: \RR^K \to \RR^K$ with maximum degree $g=3$. We can write $F(\B x) = \B A\B x + \B B \B x^{\otimes 2} + \B C \B x^{\otimes 3}$ using eq. (\ref{kronpoly}). 
Naively, a network of neurons that approximates the solution to $\DB x = F(\B x)$ can be written using eq. (\ref{eq_nonlindyn}) as
\[
\DB v = -\lambda \B v + \B \Omega_f \B s + \B \Omega_s^{m1} \B r + \B \Omega_s^{m2} \B r^{ \otimes 2} + \B \Omega_s^{m3}  \B r^{\otimes 3},
\]
which contains the third-order synapses in the last term.
However, we now reintroduce the first network (from Section 4.1) which takes $\B D \B r$ as input and outputs $W {\bm\rho} \approx \HB x ^{\otimes 2}$. This allows the term $\B \Omega_s^{m3}  \B r^{\otimes 3}$ to be replaced by $\B D\tran C (\B D \otimes \B W) (\B r \otimes \B {\bm\rho})$, yielding
\begin{equation}\label{eq_nonlin_notripl}
\begin{split}
    \DB v &= -\lambda \B v + \B \Omega_f \B s + \B \Omega_s^{m1} \B r + \B \Omega_s^{m2} \B r^{\otimes 2} + \B \Omega_s^{ext} (\B r \otimes \B {\bm\rho}) ,
\end{split}    
\end{equation}
where $\B \Omega_s^{ext} = \B D\tran \B C (\B D \otimes \B W)$.
The same argument can be extended to higher order multiplications, at the cost of having $\lceil \log_2(g) - 1 \rceil$  support networks, where $g$ is the maximum degree of $F()$ \footnote{The maximum multiplication order that can be computed with $k$ networks is $2^k$ - in fact, the first support network can compute $\hat{x}^{\thinkron 2}$, the second can take as input the output of the first and compute $\hat{x}^{\thinkron 4}$, and so on.}.

\subsection{Example: approximating a double pendulum}

\begin{figure}
\centering
\includegraphics[width=1.07\textwidth]{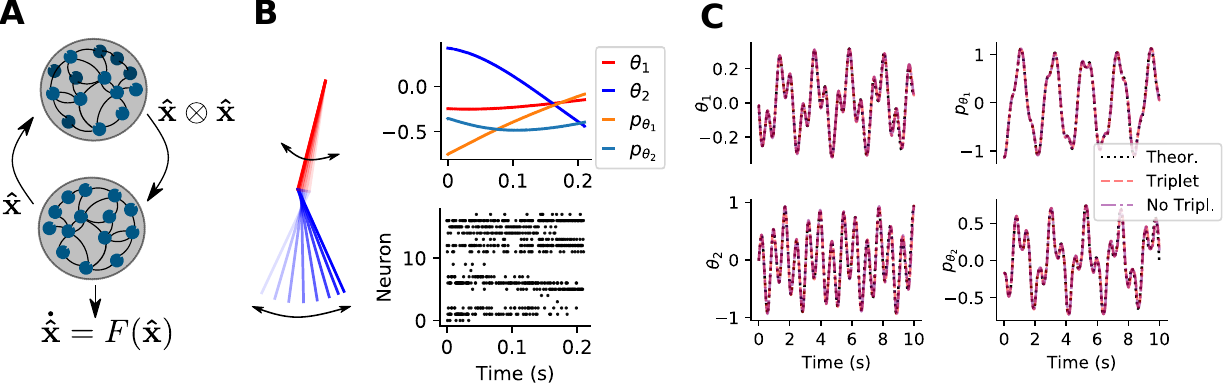}
\caption{Third order polynomial dynamics solved by sequential pairwise multiplications. 
\textbf{(A)} To avoid third order multiplications, another network can be used whose output will be the pairwise multiplication of any two input dimensions (which can be done through only pair-wise synapses). 
\textbf{(B)} Example output of a network computing the double pendulum and using an external network to avoid the third order multiplications.
\textbf{(C)} Solution computed by employing a neural network with third order synapses (dashed line) or employing two neural networks to avoid the third order multiplications (dash-dotted line) compared to the numerical solution of dynamical system (dotted line). All solutions almost perfectly overlap.}
\label{Fig4}
\end{figure}

We illustrate the use of the higher-order polynomial implementation using the double pendulum as an example (Fig.\ \ref{Fig4}B). Suppose that each pendulum has length $l$ and mass $m$. We denote $\theta_1, \theta_2$ the angles of the first and second pendulum with respect to the vertical axis (i.e. $\theta_i = 0$ when the pendulum is pointing downwards), and $p_{\theta_1}$ and $p_{\theta_2}$ their momenta. The full double pendulum dynamics can be derived using the Lagrangian (Methods~\ref{Methods:FirstOrderPendulum}; e.g., \parencite{levien1993double}). For small angles one can consider the approximation $\sin \theta \approx \theta$ and $\cos \theta \approx 1$, which leads to the following approximated dynamics:

\begin{align*}
{\dot \theta_1} &= \frac{6}{7ml^2} \left( 2 p_{\theta_1} - 3 p_{\theta_2} \right) \\
{\dot \theta_2} &= \frac{6}{7ml^2} \left( 8 p_{\theta_2} - 3 p_{\theta_1} \right) \\
{\dot p_{\theta_1}} &= -\tfrac{1}{2} m l^2 \left ( {\dot \theta_1} {\dot \theta_2} (\theta_1-\theta_2) + 3 \frac{g}{l} \theta_1 \right ) \\
{\dot p_{\theta_2}} &=  -\tfrac{1}{2} m l^2 \left ( -{\dot \theta_1} {\dot \theta_2} (\theta_1-\theta_2) + \frac{g}{l} \theta_2 \right ).
\end{align*}

We denote $\B x = (\theta_1, \theta_2, p_{\theta_1}, p_{\theta_2})\tran$, and rewrite the system as $\DB x = \B A \B x + \B C \B x^{\otimes 3}$ ($\B A$ and $\B C$ defined explicitly in Methods~\ref{Methods:FirstOrderPendulum}). \\

We implemented the first-order approximated double pendulum system in three distinct ways. As before for the Lorenz system, we first simulated a control network that was simply asked to autoencode the dynamics directly, which were computed externally. Next, we implemented two mSCNs --- one network computed the dynamics through explicit third-order multiplicative synapses (as in eq.~(\ref{eq_nonlindyn})). The other implementation utilized a support network (as explained in eq.~(\ref{eq_nonlin_notripl})), allowing the main network to avoid explicit third-order interactions. We found that these two mSCN implementations produced accurate representations of the dynamics, closely following the autoencoder network which received the ``true'' solution directly.

%%%%%%% Suggestion fig 4 %%%%%%%%%%%

%\footnote{SWK: Comments on Figure 4. Panel a+b: should use the same conventions as figure 1 for multplicative synapses interacting. Panel  c: can simplify the legend description. Can add two arrows illustrating motion. Panel d: I think we can leave this out possibly? Or combined with b. }

\section{On the number of required connections}
As shown in the previous sections, mSCNs offer a powerful and intuitive way of implementing polynomial dynamical systems in spiking networks. However, though they may be efficient with respect to the number of neurons and spikes required, they can require dense synaptic connections (sometimes with several connections for each pair of neurons). 
In the standard SCN framework, any two neurons
can be connected by fast and slow synapses. In mSCNs, additional connections are introduced with the multiplicative synapses. So far we have assumed full all-to-all connectivity of all types of connections (i.e., $N^2$ fast connections, $N^2$ slow connections, $N^3$ pair-wise multiplicative synapses, and so on). However, connectivity in the brain is known to be sparse \parencite{song2005highly, lefort2009excitatory}. Given this constraint it is important to understand
how mSCNs can be constructed with sparser connectivity, and
the relationship between connectivity density and performance. \\

Consider networks of $N$ neurons representing a $K$-dimensional signal space, with decoding matrix $\B D\in\mathbb{R}^{K\times N}$. The $i-$th column vector of the matrix $\B D$, denoted by $\B D_i$, represents the weights associated to neuron $i$. Here and in the following we will say that ``neuron $i$ codes for dimension $x$'' meaning that the $x-$th entry of $\B D_i$ is $\neq 0$. 
%Any given unit in our setting could participate in the coding of an arbitrary number of dimensions, given by the number of nonzero terms of the decoding weights $\B D_i$.  
We will define the connectivity density as the proportion of the all-to-all connectivity which is being used. Connectivity density is then determined by the decoder matrix $\B D$ and some fixed matrices $\B A_d$ given by the dynamical system, as explained in eq.~(\ref{eq_nonlindyn}). Thus far, 
%%% 11.09.2021 - changed after second round of review %%%
% we have considered that each neuron codes for each dimension, 
we have considered that each neuron codes for all dimensions, 
%%% end changes %%%
meaning that $\B D$ is dense and connectivity is all-to-all for each synapse type. If neurons instead only coded for a subset of the dimensions, how sparse would the connectivity be? We will make this explicit by giving each neuron a fixed probability $p$ to code for each signal dimension (i.e. $p$ is the probability that a given matrix entry in $\B D$ is non-zero). If $p=1$, then all neurons will code for all signal dimensions. As $p$ approaches zero, neurons will code for progressively less dimensions. \\

\begin{figure}
\centering
\includegraphics[width=0.99\textwidth]{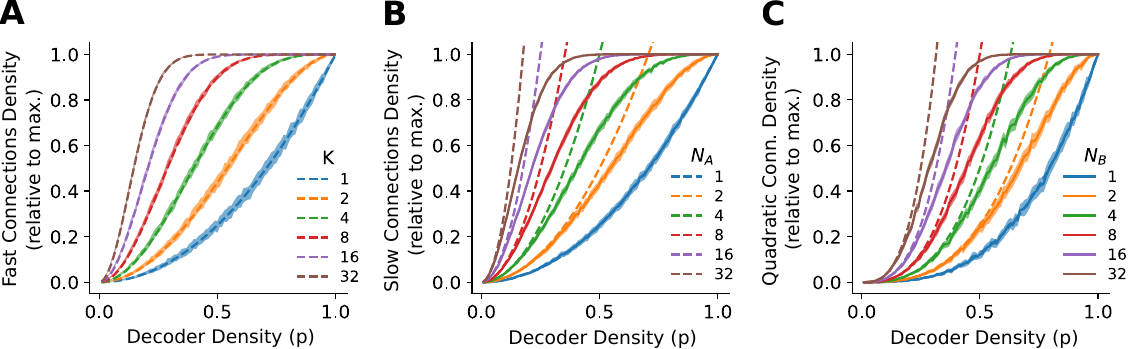}
\caption{Expected connectivity density relative to all-to-all connectivity for the different synapse types  as a function of the decoder density $p$. Solid lines illustrate mean across iterations, whereas shaded areas represent $\pm 1$ standard deviation from the average of the simulated connectivity (see Methods~\ref{Methods:Connectivites}). Dashed lines correspond to theoretical upper bounds (Eq.~\ref{Eq:upperbound1}-\ref{Eq:upperbound3}). (\textbf{A}) The connectivity density for the fast synapses for different signal dimensions ($K$). (\textbf{B}) The connectivity density for the slow synapses due to linear computation (i.e. $\sim N_A p^2$), for different linear dynamical system densities ($N_A$). (\textbf{C}) The connectivity density for slow multiplicative (quadratic) synapses, for different quadratic dynamical system densities ($N_B$). $N_A$ and $N_B$ represent the number of non-zero entries of matrices $\B A$ and $\B B$ respectively. }
\label{Fig3}
\end{figure}

Consider the three connectivity matrices required for second-order multiplicative computations: the fast connections $\B \Omega_f=\B D\tran \B D$, the slow connections $\B \Omega^{m1}_s = \B D\tran (\B A + \lambda \B I) \B D$ and the multiplicative connections $\B \Omega^{m2}_s = \B D\tran \B B \B D\thinkron \B D$. In Fig.~\ref{Fig3} we investigate the relationship between decoder density and final network connectivity density. We calculate theoretical upper-bounds on the expected connectivity and report here the asymptotic behavior in terms of density (see Methods \ref{Methods:Connectivites} for detailed derivations). Additionally, by generating random decoding matrices for different probabilities $p$, we measure the empirical expected connectivity density. %%The fast connections only depend on the density of the decoder, and any two neurons are connected whenever they share a decoding-dimension. For the slow and multiplicative connections this density additionally depends on the `density' of the dynamical system interactions, i.e. the number of non-zero elements in the matrices $\B A$ and $\B B$, denoted with $N_A$ and $N_B$ respectively.  Although this connection probability rises sharply as the signal dimensionality increases, we see that for low to moderate $p$ and $K$ the connectivity density is far below all-to-all.
The fast connection density (defined as the proportion of maximum number of connections used) depends on both $p$ and the signal dimension $K$ and behaves as
 \begin{equation}\label{Eq:upperbound1}
 \mathbb{E}(\texttt{fast conn. dens.}) \sim  1 - (1-p^2)^K.
 \end{equation}
Although the density rises sharply as the signal dimensionality increases (Fig.~\ref{Fig3}A), we see that for low to moderate $p$ and $K$ the connectivity density is far below all-to-all. 
 The slow and multiplicative connection densities additionally depend on the `density' of the dynamical system interactions, i.e. the number of non-zero elements in the matrices $\B A$ and $\B B$, denoted with $N_A$ and $N_B$ respectively.
 We find theoretical upper bounds on the expected densities
 \begin{equation}\label{Eq:upperbound2}
 \mathbb{E}( \texttt{slow conn. dens.}) \lesssim \mathbb{E}(\texttt{fast conn. dens.}) + N_A  p^2,
 \end{equation}
 for slow connections (Fig.~\ref{Fig3}B) and
 \begin{equation}\label{Eq:upperbound3}
 \mathbb{E}(\texttt{multiplicative conn. dens.}) \lesssim N_B p^3,
 \end{equation}
 for multiplicative connections (Fig.~\ref{Fig3}C, see Methods~\ref{Methods:Connectivites} for details).  Notably, the multiplicative connections have a much slower rise compared to the fast connections, which may push the network closer to a biologically-plausible regime.

Such reduced connectivity density does come at some cost to performance, in particular in robustness. We demonstrate this by considering the Lorenz attractor implementation with different values for $p$, and by considering different numbers of active neurons. We found that decreasing the decoder density made the network more prone to errors as the number of lost cells increased (Supp.~Fig.~\ref{SupFigRobust}C+D). Nonetheless, this increase in error was nearly nonexistent and became sizeable only after killing more than 80\% of the cells.

%%%%%%%% insert figure 3 here %%%%%%%%%%%
%%%% end figure 3 %%%%%%%%%%%%%%

\section{Discussion}
% P1 -- summary of work
In this report, we investigated a new approach for implementing nonlinear dynamical systems in spiking neural networks. We extended the spike coding network (SCN) framework \parencite{boerlin2013predictive, deneve2016efficient, calaim2020robust} to implement arbitrary polynomial dynamics. We obtain the multiplicative SCN (mSCN), which requires fast, slow, and multiplicative connections. For second-order systems, the connectivity requires pair-wise multiplicative synapses, and we demonstrated how higher-order multiplications can be implemented in several network stages with only pair-wise synapses. We demonstrated the accuracy and flexibility of this formalism by simulating the Lorenz system and a third order approximation of a double pendulum. Due to the rich flexibility of polynomials for approximating arbitrary nonlinear functions (Stone-Weierstrass Theorem or Taylor Expansion), this approach could in principle be extended to many other systems for which a lower-order polynomial approximation is sufficient. Lastly, we analyzed the relationship between the sparsity of signal coding per neuron and the connectivity sparsity (for fast, slow, and multiplicative connections), and showed how the need for all-to-all connectivity can be relaxed.

\subsection{Related work}
The study of nonlinear computation has a long history in computational neuroscience, and has traditionally been studied in firing-rate networks, in which each neuron is represented by a continuous (or sometimes binary) variable \parencite{dayan2001theoretical}. When such neurons are endowed with a nonlinear input-output function, nonlinear computations are possible \parencite{jaeger2001echo, sussillo2009generating, mante2013context, rubin2015stabilized, mastrogiuseppe2018linking}. \\

In spiking neural networks (SNNs) nonlinear computations have previously been achieved by using random connectivity as a basis for complex dynamics \parencite{maass2002real}, or using supervised training algorithms to optimize the networks \parencite{neftci2019surrogate}. Alternatively, more principled approaches to building spiking neural networks include the neural engineering framework (NEF) \parencite{eliasmith2004neural} and SCNs \parencite{boerlin2013predictive}, the latter of which we studied here. Nonlinear dynamical system implementations in previous SNNs usually harnessed various nonlinearities (e.g. neural, dendritic, or synaptic) as basis functions, with the connectivity required for a given dynamical system implemented through supervised training (as we also implemented in Fig.~\ref{Fig2}C). Previous nonlinear computations in SCNs also relied on such basis functions \parencite{thalmeier2016learning, alemi2018learning}. The basis function approach works very well, but does not allow for a direct derivation of the connectivity given a nonlinear dynamical system and thus lacks a well-defined mapping between computation and connectivity. In contrast, in mSCNs the connectivity is defined directly from the dynamical system, and is therefore more interpretable. \\

Finally, a more recent approach has emerged in which spiking networks are treated as piecewise linear functions, depending on which neurons in the network are active \parencite{mancoo2020understanding, baker2020nonlinear} (which may also explain the success of supervised training algorithms without explicit basis functions, e.g., \parencite{zenke2021remarkable}). These methods can be seen as complementary to our approach, in which case a nonlinear input-output function would be combined with nonlinear dynamics through additional slow connections.

\subsection{Biological implications}
mSCNs represent a novel hypothesis of nonlinear computation in neural circuits, which can be seen as complementary to previous basis-function approaches. While previous theoretical studies have noted the wide range of nonlinear computations that multiplicative synapses could in principle enable \parencite{koch_multiplying_1992, salinas1996model, nezis_accurate_2011},  our work provides a precise formalization on how to use such synapses to implement any polynomial dynamical system efficiently in a spiking neural network. Additionally, the resulting networks inherit the attractive features from standard SCNs which match well to biological network activity (e.g. irregular and sparse activity, robustness, and E/I balance). But how biologically feasible are the the required synapse interactions and network connectivities? \\

First, the resulting network connectivity is extremely dense: by default, mSCN networks predict all-to-all connectivity. While local circuits in the cortex are indeed densely connected \parencite{fino2011dense, harris2013cortical}, connectivity is not all-to-all \parencite{ko2011functional,song2005highly, perin2011synaptic}. However, as we showed in Fig.~\ref{Fig3}, full all-to-all connectivity is not required for mSCNs to function. The connectivity density is a function of the fraction of variables each neuron codes for, which is formalized by imposing sparsity of the decoder matrix. Consequently, according to SCN theory, if all neurons would code for all dimensions, this would be reflected in an all-to-all connectivity structure. Conversely, if neurons would code for a subset of the dimensions, this would be reflected in a less dense connectivity structure (but also, accordingly, a less robust code). This last property is not easily measured in the brain, where the amount of variables each neuron codes for is not clear. While current hypotheses emphasize that neurons exhibit mixed selectivity to many task variables \parencite{rigotti2013importance}, recent evidence also suggests that cortical representations are very high-dimensional \parencite{stringer2019spontaneous,stringer2019high}. Therefore, it is plausible that neurons can display mixed selectivity while also only coding for a small subset of the overall coding space, leaving room for a more sparse implementation of our model. \\

Second, mSCNs require each pair of neurons to have three types of synapses connecting to each other (Fig.~\ref{Fig1}). The presence of several distinct synaptic connections between pairs of cells have indeed been observed in experimental circuit reconstruction studies \parencite{popov2009complexity,kasthuri2015saturated}. However, the feasibility of the precise multiplicative interactions is less clear. We can at the very least state that a form of multiplication must be performed at some level in biological networks. Several examples of effectively multiplicative computations have been characterized from experimental studies \parencite{pena_auditory_2001, gabbiani_multiplicative_2002, zhou_multiplicative_2007,arandia2016multiplicative}, and multiplicative interactions have been hypothesized to exist in the dendritic tree \parencite{london2005dendritic}. Indeed, experimental and theoretical work has long shown the computational advantages of nonlinear synaptic and dendritic interactions in single neurons \parencite{poirazi2003pyramidal, london2005dendritic}. Mechanisms such as dendritic calcium or NMDA spikes \parencite{schiller2000nmda, augustine2003local}, synaptic clustering \parencite{larkum2008synaptic}, and shunting inhibition \parencite{mitchell2003shunting, zhang2013nonlinear} are well established and could contribute to a code relying on multiplicative interactions (as detailed in \parencite{koch_multiplying_1992}). Finally, even if not fully feasible, for a given polynomial system mSCNs could be seen as defining the \textit{ideal} set of interactions required for a given network of neurons, which would then have to be replicated by a given network either by a basis function approach or through non-optimal multiplicative interactions. Future work will have to show how well a more biological implementation of multiplicative interactions would allow for precise nonlinear computations. \\

Finally, while higher order multiplicative interactions are increasingly difficult to implement biologically, we demonstrated that by stacking networks with lower-order interactions one can achieve the same computations. This makes the concrete prediction that the connectivity between areas should be of the same dimensionality as the signal being transferred. This does indeed seem to be the case to some degree, with the communication between some areas being low-dimensional \parencite{semedo2019cortical}. There is also a likely limit to how many networks can be effectively stacked in this way to perform higher-order interactions, as each stacked network introduces a delay.

\subsection{Computational and neuromorphic applications}
Even given potential biological limitations, we contend that mSCNs offer two benefits. First, the fact that the implementation of polynomial dynamics is direct and explicit implies that this technique offers a useful comparative control when considering the possible computations that spiking networks can perform, as well as the limits of their accuracy. This model may therefore serve as a useful reference for future studies. Second, neuromorphic implementations of spiking neurons are becoming increasingly feasible \parencite{young_review_2019}. As many of these rely on networks of integrate-and-fire type neurons \parencite{indiveri2011neuromorphic, merolla2014million, davies2018loihi}, it is in principle less constrained by biological plausibility. In this context, the need for dense pair-wise connectivity and precise multiplicative interactions may be an acceptable cost for a precise connectivity recipe for a given nonlinear computation with fewer neurons. For example, a Lorenz attractor can be efficiently and precisely implemented with just 10 neurons (Supp.~Fig.~\ref{SupFigRobust}), in contrast to the 1600 neurons used in \parencite{thalmeier2016learning}.

\subsection{Conclusion}
In sum, we have provided a proof of concept of direct and explicit polynomial dynamics implemented in spiking networks. Future directions include the application of this framework to other biologically-plausible and neuromorphic computations, a study of the efficiency of this framework, and the potential for biologically-plausible learning of the connectivity.

\section*{Author Contributions}
MN contributed to discussions, initial and follow-up implementations, mathematical derivations and writing of the paper; JWP contributed to discussions and initial implementations; WFP contributed to discussions, supervision, and writing of the paper; SWK conceived the initial idea, contributed to supervision, discussions and initial implementation, and writing of the paper.

\section*{Supplementary material}
Script and codes are available online: \url{https://github.com/michnard/mult_synapses}

\section{Methods}

\subsection{General derivation of spike coding network}\label{Methods:GeneralDerivation}

We will show here
%the derivation of the connectivity and dynamics of a network that optimally represents a modification of the incoming signal. This is
a generalization of the derivation of spike-coding networks (SCNs) shown in \cite{barrett2016optimal}, ignoring the constraint that neurons need to be either excitatory or inhibitory.
Consider a network of $N$ leaky integrate-and-fire neurons receiving time-varying inputs $\B{x}(t) = (x_1(t), \dots, x_K(t))$, where $K$  is the dimension of the input. 
%In response to this input, the neurons in the network will emit spikes. 
For each neuron $i$ we denote with $s_i(t) = \sum_k \delta(t^i_k - t)$ the spike train function, where $\delta$ represents the Dirac delta function and $\{t^i_k \geq 0 \}$ is the set of discrete times at which a spike was emitted. 
The population spike train function is described by the vector $\B s(t) = (s_1(t), \dots, s_N(t))\tran$. 
We define the filtered spike trains (loosely called firing rate) of neuron $i$ as a convolution of the spike train with an exponentially decaying kernel
\begin{align}\label{exp_dec_kern}
    r_i(t)  = \int_0^t \exp(-\lambda{t'}) s_i(t - t') dt' = \sum_{t^i_k \leq t} \exp(-\lambda(t - t^i_k)))
\end{align}
with leak constant $\lambda$, or, equivalently, in the differential form
\[ 
\dot{r_i}(t) = - \lambda{r_i(t)} + s_i(t).
\]
We denote the firing rate for all neurons by the vector $\B r(t) = (r_1(t),\dots,r_N(t))\tran$.
Vectors will be denoted by bold letters, and wherever possible we will exclude the explicit dependence on time for the sake of text clarity.\footnote{Throughout the text, the input signals, the membrane voltages and the spike trains are all time-dependent quantities, whereas the thresholds, the decay constants, and the connection strengths are all constants.}
A neuron $i$ fires a spike whenever its membrane potential, $V_i$ exceeds a spiking threshold, $T_i$, and is then reset to the value $V_i=R_i$. 
%We can write this as $V_i>T_i$, where $V_i$ is the membrane potential of neuron $i$ and $T_i$ is the spiking threshold.
 %When a neuron spikes, its membrane potential is reset to the value $V = R_i$. %For ease of presentation, we include this reset in the self-connections. 

Consider a generic smooth function $ G : \RR^K \to \RR^M$, $M\geq 1$. Our goal is to derive dynamics and connectivity of the network so that its output activity provides an accurate representation of the  modification of the incoming signal $\B y = G(\B x) \in \RR^M$. Notice that, using the identity function, one can recover the same form considered in \cite{barrett2016optimal}.

Following the assumptions made in the main text, we require the signals to be linearly decodable, so that the readout can be simply written as $\hat{ \B y} = \B D \B r \approx G(\B x)$. The matrix $\B D \in \RR^{M \times N}$ is called the decoding matrix, and its $i$-th column vector  $\B D_i \in \RR^M$ is the fixed contribution of neuron $i$ to the signal.
The accuracy of the representation is measured using a squared error loss function, $E = \Vert\B y - \hat{\B y}\Vert^2_2 = \Vert G(\B x) - \hat{\B y}\Vert^2_2$. The second assumption made in the main text requests the network to be efficient, and can be formalized by asking that a neuron fires a spike only if its effect on the readout will reduce the loss function:
\[
E(\text{spike}) < E(\text{no spike}),
\]
which is, noticing that a spike of neuron $i$ changes the readout by $\HB y \to \HB y + \B D_i$,
\begin{align}
    \left\Vert G(\B x) - (\HB y + \B D_i)\right\Vert^2_2 < \left\Vert G(\B x) - \hat{\B y}\right\Vert^2_2.
\end{align}
After expanding the squares and canceling equal terms we obtain
\begin{align}
    \left\Vert\B D_i\right\Vert^2_2 - 2\B D_i \tran (G(\B x) - \hat{\B y}) < 0,
\end{align}
which can be rearranged into 
\begin{align}
     \B D_i \tran (G(\B x) - \hat{\B y}) > \frac{\left\Vert\B D_i\right\Vert^2_2}{2}.
\end{align}
This equation is crucial: it describes a spiking rule under which the loss function is reduced, and it offers an enticing geometric interpretation of the behavior of the network \citep{calaim2020robust}.
The right hand side of the equation is fixed, and can be interpreted as the spiking threshold of neuron $i$: 
\[
T_i = \frac{\left\Vert\B D_i\right\Vert^2_2}{2}.
\]
The left hand side of the equation, similarly to the derivation showed in \citep{barrett2016optimal}, is used to define the voltage of neuron $i$
\begin{equation}\label{V_i_G}
    V_i = \B D_i\tran(G(\B x) - \HB{y}),
\end{equation}
which, taking the derivative, yields,
\begin{equation}\label{V_i_prime}
\begin{split}
    \dot V_i &= \B D_i\tran\left(\frac{dG(\B x)}{dt} - \frac{d\HB{y}}{dt} \right) \\
    &=\B D_i\tran (\B{J}_G(\B x) \dot{\B x}) + \B D_i\tran \lambda \HB{y} - \sum_k \B D_i\tran \B D_k s_k,
\end{split}
\end{equation}
where we used $\B{J}_G$ to indicate the Jacobian of the function $G$.
Using (\ref{V_i_G}), we have that $\B D_i \tran \HB{y} = -V_i + \B D_i\tran(G(\B x)$, and substituting this into (\ref{V_i_prime}) we obtain:
\begin{equation}\label{Nonlin_repr}
    \dot V_i = -\lambda V_i
            + \B D_i\tran\left( \B{J}_G(\B x) \dot{\B x} + \lambda G(\B x)\right)
            - \sum_k \B D_i\tran \B D_k s_k.
\end{equation}
This equation describes the dynamic behavior of the voltage of a neuron in a network that represents $G(\B x)$. We will use the vector form 
\begin{equation}\label{Nonlin_repr_vec_meth}
    \DB v = -\lambda \B v
            +  \B D\tran\left( \B{J}_G(\B x) \dot{\B x} + \lambda G(\B x)\right)
            + \B \Omega_f \B s,
\end{equation}
where $\B \Omega_f = - \B D\tran \B D$ represents the fast connections among units, and also includes the reset terms on the diagonal.

\subsection{The Kronecker product\label{Methods:kronecker}}
Throughout the text we make heavy use of the Kronecker product. $\otimes$ represents the Kronecker product, which is defined for any couple of matrices $\B A, \B B$ of any arbitrary size as 
\[
\B A\otimes \B B = 
\begin{bmatrix}
  a_{11} \B{B} & \cdots & a_{1n}\B{B} \\
             \vdots & \ddots &           \vdots \\
  a_{m1} \B{B} & \cdots & a_{mn} \B{B}
\end{bmatrix}.
\] 
We often use the mixed-product property, which states: If $\B A, \B B, \B C$ and $\B D$ are matrices of such size that one can form the matrix products $\B A \B C$ and $\B B \B D$, then
\[(\B A \otimes \B B)(\B C\otimes \B D) = (\B A\B C) \otimes (\B B\B D).\] 

\subsection{Representation of the multiplication of incoming inputs}\label{Methods:RepresentingMultiplication}

Consider the function $G: \RR^K \to \RR^{K^2}$ defined as $G(\B x) = \B x \otimes \B x$, where $\otimes$ is the Kronecker product.
Suppose that the input $\B x = (x_1, \dots ,x_K)$ is given as a linearly decodable input from an upstream network, such that $\B x = \B D \B r$, where $\B r$ describes the filtered spike-trains of the upstream neurons and $\B D$ is their decoding matrix. 
This generalization requires us to keep track of $\DB x$: if the upstream neurons follow equation (\ref{exp_dec_kern}), and we denote with $\B s$ their spike trains, we will have that 
%$\DB x = \B W \DB {\bm\rho} = \B W ({\bm\sigma} - \alpha {\bm\rho})$
$\DB x = \B D \DB r = \B D (\B s - \lambda \B r)$
, where $\lambda$ represents their leak constant. In order to use eq.  (\ref{Nonlin_repr_vec_meth}), we need to compute the Jacobian of the function $G$. That's given by the $K^2 \times K$ matrix $\B J_G(\B x)$, with column $i$ given by $\frac{dG}{dx_i}$. 
Denote with $\B D^i$ the $i-$th row of the matrix $\B D$, and with $\left[ \B{J}_G(\B x) \DB x \right]_{ iK + j}$ the $(iK + j)-$th entry of the matrix-vector product $\B{J}_G(\B x) \DB x \in \RR^{K^2}$, for $0<i \leq j \leq K$. We have:
\begin{equation*}
\begin{split}
\left[ \B{J}_G(\B x) \DB x \right]_{ iK + j} 
    &= \frac{dG}{dx_i} \frac{dx_i}{dt} + \frac{dG}{dx_j} \frac{dx_j}{dt} \\
    &= x_j \dot x_i + x_i \dot x_j \\
    &= (\B D^j \B r)(\B D^i (\B s - \lambda \B r)) + (\B D^i \B r)(\B D^j (\B s - \lambda \B r) \\
    &= (\B D^j \otimes \B D^i + \B D^i \otimes \B D^j)(\B r \otimes (\B s - \lambda\B r)) \\
    &= (\B D^i \otimes \B D^j)(\B r \otimes \B s + \B s \otimes \B r - 2 \lambda \B r \otimes \B r)
\end{split}
\end{equation*}
and
\[
\left [G(x) \right]_{iK + j}  = (\B D^i \B r)(\B D^j \B r) = (\B D^i\otimes \B D^j)(\B r \otimes \B r).
\]
We can now derive the voltage equations of a network of neurons that represents the product of any pair of input dimensions using the equation derived in the previous section. Denote with ${\bm\sigma}$ the spike train of the network, and with ${\bm\rho}$ their filtered spike train with leak constant $\alpha$. Using eq. (\ref{Nonlin_repr_vec_meth}) we have
\[
\DB v = -\lambda \B v  + \B \Omega_{x} (\B r \otimes \B s + \B s \otimes \B r + (\alpha - 2 \lambda) \B r \otimes \B r) + \B \Omega_f^W \B {\bm\sigma},
\]
with $\B \Omega_x = \B W\tran (\B D \otimes \B D)$, $\B \Omega_f^W = - \B W \tran \B W$ and $\B W$ being their decoding matrix. An example of the output of such a network can be seen in Supp. Fig. \ref{SupFigKronProd}. In that case the input was $3-$dimensional, and the $9-$dimensional output faithfully represented the product of each input dimension pair.
% \[
% \DB v = -\lambda \B v  + \Omega_{x} ({\bm\rho} \otimes {\bm\sigma} + {\bm\sigma} \otimes {\bm\rho} + (\lambda - 2 \alpha) {\bm\rho} \otimes {\bm\rho}) + \B \Omega_f \B s
% \]
% where $\Omega_f =  \B D\tran \B D$ and $\Omega_x = \B D\tran (\B W \otimes \B W)$.

\subsection{Implementing dynamical systems in spike coding networks}\label{Methods:LinearNonlinearSCNS}

By using the identity function $G(\B x) = \B x$ in (\ref{Nonlin_repr_vec_meth}) we obtain the ``classical'' equation

\begin{equation} \label{eq_repr_meth}
\dot{\B v} = -\lambda \B v + \B D\tran(\dot{\B x} + \lambda \B x) + \B \Omega_f \B s.
\end{equation}
This will be the starting point to implement linear and nonlinear dynamical systems. Linear dynamical systems were already considered in \citep{boerlin2013predictive}. Here we will focus on a more general class of nonlinearities, namely polynomial nonlinearities, and show that the original formulation can be analytically extended to implement any polynomial nonlinearity.

Denote with $F: \RR^K \to \RR^K$ the dynamic under study, so that $\dot{\B x} = F(\B x)$. Starting from (\ref{eq_repr_meth}) and knowing that $\B x \approx \hat{\B x}$ we can consider the following approximation:
\begin{equation} \label{eq2_meth} 
\dot{\B v} = -\lambda  \B v + \B D\tran(F(\hat{\B x}) + \lambda  \hat{\B x}) + \B \Omega_f \B s.
\end{equation}
If $F$ is a linear dynamic of the form $F(\B x) = \B A \B x$, with the matrix $\B A \in \RR^{K\times K}$, we recover the same form considered in \citep{boerlin2013predictive}:
\begin{equation} \label{eq_lindyn_meth}
\begin{split}
\dot{\B v}  &= -\lambda  \B v + \B D\tran\left(\B A \hat{\B x} + \lambda  \hat{\B x}\right) + \B \Omega_f \B s \\
            &= -\lambda  \B v + \B D\tran\left(\B A \B D \B r + \lambda \B \B D \B r\right) + \B \Omega_f \B s \\
            &= -\lambda  \B v + \B \Omega_s \B r + \B \Omega_f \B s ,
\end{split}
\end{equation}
where $\B \Omega_f = - \B D\tran \B D$ and $\B \Omega_s = \B D\tran (\B A + \lambda \B I)\B D$ represent the fast and slow connections respectively. 

If $F$ is polynomial, we proceed as follows. Using Kronecker notation, any polynomial $F:R^K \to R^K$ with maximum degree $g$ can be written in the form
\begin{equation}\label{kronpoly_meth}
    % F(\B x) = \sum_{d=0}^g A_d \otimes^d(\B x)
    F(\B x) = \sum_{d=0}^g \B A_d \B x^{\otimes d},
\end{equation}
where $\B A_d \in \RR^{K \times K^d}$ is the matrix of coefficients for the polynomials of degree $d$, and we define $\B M^{\otimes d} = \B M \thinkron \B M \thinkron \cdots \thinkron \B M$ as the Kronecker product applied $d$ times, with the convention that $\B M^{\otimes 0} =
1$ and $\B M^{\otimes 1} = \B M$.
Once again replacing $\B x$ by $\hat{\B x}$, as well as using the notation introduced in (\ref{kronpoly_meth}) and the mixed-product property, we get
\begin{equation*}% \label{F_hat_x}
\begin{split}
F(\HB x) &= \sum_{d=0}^g \B A_d \HB x^{\otimes d} \\
              &= \B A_0 + \B A_1 \HB x + \B A_2 \HB x ^{\otimes 2} + \B A_3 \HB x^{\otimes 3} + \dots \\
              &= \B A_0 + \B A_1 \B D \B r + \B A_2 (\B D\B r)^{\otimes 2} + \B A_3 (\B D\B r)^{\otimes 3} + \dots \\
              &= \B A_0 + \B A_1 \B D \B r + \B A_2 (\B D^{\otimes 2})(\B r ^{\otimes 2}) + \B A_3 (\B D^{\otimes 3})(\B r^{\otimes 3}) + \dots \\
              &= \sum_{d=0}^g \B A_d \B D^{\otimes d} \B r^{\otimes d} .
\end{split}
\end{equation*}

Inserting it into (\ref{eq2_meth}) one obtains the equations describing a network of integrate-and-fire neurons that approximate the solution of a polynomial dynamical system:
\begin{equation} \label{eq_nonlindyn_meth} 
\begin{split}
\dot{\B v} &= -\lambda \B v - \B D\tran \B D \B s + \B D\tran(\sum_{d=0}^g \B A_d \B D^{\otimes d} \B r^{\otimes d} + \lambda \hat{\B x}) \\
      &= -\lambda \B v + \B \Omega_f \B s + \B \Omega^{m0}_s + \B \Omega^{m1}_s \B r + \B \Omega^{m2}_s \B r^{\otimes 2} + \dots + \B \Omega^{mg}_s \B r^{\otimes d} \\
      &= -\lambda \B v + \B \Omega_f \B s + \sum_{d=0}^g \B \Omega^{md}_s \B r^{\otimes d},
\end{split}
\end{equation}
where $\B \Omega_f = - \B D\tran \B D$, $\B \Omega_s^{m1} = \B D\tran (\B A_1 + \lambda \B I)\B D$ and $\B \Omega^{md}_{s} = \B D\tran \B A_d \B D^{\otimes d}$ for $d \in \{0,2,3,\dots,g\}$.

\subsection{Implementing the Lorenz system}\label{Methods:Lorenz}
Denoting $\B x = (x,y,z)\tran$, the Lorenz attractor can be described in the form of eq.~\ref{kronpoly_meth} as
\[
\dot{\B x} = \B A  \B x + \B B \B x ^{\otimes 2},
\]
where 
\begin{equation}\label{matrices}
\B A = \begin{bmatrix}
-\sigma & \sigma & 0 \\
\rho & -1 & 0 \\
0 & 0 & -\beta
\end{bmatrix},
\end{equation}
and $\B B \in \RR^{3\times9}$ with $B_{23}=-1$, $B_{32}=1$, and all other elements of $\B B$ being zero.

Following eq.~\ref{eq_nonlindyn_meth} the corresponding voltage dynamics in an mSCN are described by
\[
\dot{\B v} = -\lambda \B v + \B \Omega_f \B s + \B \Omega^{m1}_s \B r + \B \Omega^{m2}_s \B r^{\otimes 2},
\]
where $\B \Omega_f = - \B D\tran \B D$, $\B \Omega^{m1}_s = \B D\tran (\B A + \lambda \B I)\B D$ and $\B \Omega^{m2}_s = \B D\tran \B B \B D ^{\otimes 2}$.

\subsection{Learning nonlinear dynamics through basis functions}\label{Methods:BasisFuns}
In previous work the standard SCN derivation was extended to implement arbitrary nonlinear dynamical systems through weighted basis functions, meant to model nonlinear synapses or dendrites \citep{alemi2018learning, thalmeier2016learning}. We will use a similar approach to approximate the nonlinear part of a dynamical system of the form
\begin{equation}
    \dot{\B x} = \B A \B x + F(\B x).
\end{equation}

The basis-function approach  derivation consists replacing the function $F(\B x)$ by a weighted set of $L$ basis functions $\B g(\B x) = [g_0(\B x), \dots, g_L(\B x)]$, such that $\B C \B g \approx F(\B x)$ (where $\B C \in \RR^{K \times L}$ are the required weights). Eq.~(\ref{eq2_meth}) can then be rewritten as
\begin{equation}
\dot{\B{v}} = -\lambda \B v + \B \Omega_s \B r + \B D\tran \B C \B g(\B x) - \B \Omega_f \B s.
\end{equation}
In previous work the weights $\B C$ were found through supervised local learning rules. For brevity and comparison's sake we will instead find the optimal weights through regression (following \citep{eliasmith2004neural}).

We can find the weights by solving the following optimization problem
\begin{equation}
    \texttt{min}_{\B{C}} ||F(\B{X})-\B{C}\B{G}||_2^2,
\end{equation}
where $\B X \in \RR^{K \times M} $ are $M$ sampled inputs, $\B{G} \in\RR^{L \times M}$ are the resulting basis function outputs, and  $F()$ is the target function. The ordinary least squares (OLS) solution is then
\begin{equation}
    \B{C}_\texttt{OLS} = F(\B{X})\B{G}\tran(F(\B{X})F(\B{X})\tran)^{-1}.
\end{equation}
In previous work online learning rules were used to minimize the cost \citep{alemi2018learning, thalmeier2016learning}, but as learning rules are not the focus of this paper, we used the above solution. For the basis functions we used a simple rectification function ($g(x)=[bx+c]^+$, with randomly distributed $b\in[-1, 1]$ and $c\in[-90,90]$), but many types of nonlinearities will work. 

\subsection{First order approximation of the double pendulum}\label{Methods:FirstOrderPendulum}

The equations describing the time evolution of the double pendulum with each length $l$ and mass $m$ can be derived using the Lagrangian \citep{levien1993double}. $\theta_1, \theta_2$ describe the angles of the first and second pendulum with respect to the vertical axis (i.e. $\theta_i = 0$ when the pendulum is pointing downwards). The position of the centers of mass can be written thanks to these two coordinates: assuming that the origin is at the point of suspension of the first pendulum, its center of mass will be at:
\begin{align*}
x_1 &= \frac{l}{2} \sin \theta_1, \, \,
y_1 = -\frac{l}{2} \cos \theta_1
\end{align*}
and the center of mass of the second pendulum is at
\begin{align*}
x_2 = l \left ( \sin \theta_1 + \tfrac{1}{2} \sin \theta_2 \right ), 
y_2 = -l \left ( \cos \theta_1 + \tfrac{1}{2} \cos \theta_2 \right ).
\end{align*}
The full dynamics can be described by a $4-$dimensional dynamical system representing the two angles and the two moments:
\begin{align*}
{\dot \theta_1} &= \frac{6}{ml^2} \frac{ 2 p_{\theta_1} - 3 \cos(\theta_1-\theta_2) p_{\theta_2}}{16 - 9 \cos^2(\theta_1-\theta_2)} \\
{\dot \theta_2} &= \frac{6}{ml^2} \frac{ 8 p_{\theta_2} - 3 \cos(\theta_1-\theta_2) p_{\theta_1}}{16 - 9 \cos^2(\theta_1-\theta_2)}
\end{align*}

\begin{align*}
{\dot p_{\theta_1}} &= -\tfrac{1}{2} m l^2 \left ( {\dot \theta_1} {\dot \theta_2} \sin (\theta_1-\theta_2) + 3 \frac{g}{l} \sin \theta_1 \right ) \\
{\dot p_{\theta_2}} &=  -\tfrac{1}{2} m l^2 \left ( -{\dot \theta_1} {\dot \theta_2} \sin (\theta_1-\theta_2) + \frac{g}{l} \sin \theta_2 \right ).
\end{align*}

We will use a small angle approximation of the above equations: if $\theta \approx 0$, the functions $\sin, \cos$ are well approximated by $\theta, 1$ respectively.
The introduction of this simplifying assumption turned the above equations into these:
\begin{equation}\label{eq_doubl_pend_first_ord}
\begin{split}
{\dot \theta_1} &= \frac{6}{7ml^2} \left( 2 p_{\theta_1} - 3 p_{\theta_2} \right) \\
{\dot \theta_2} &= \frac{6}{7ml^2} \left( 8 p_{\theta_2} - 3 p_{\theta_1} \right) \\
{\dot p_{\theta_1}} &= -\tfrac{1}{2} m l^2 \left ( {\dot \theta_1} {\dot \theta_2} (\theta_1-\theta_2) + 3 \frac{g}{l} \theta_1 \right ) \\
{\dot p_{\theta_2}} &=  -\tfrac{1}{2} m l^2 \left ( -{\dot \theta_1} {\dot \theta_2} (\theta_1-\theta_2) + \frac{g}{l} \theta_2 \right ).
\end{split}
\end{equation}

These can be implemented using either equation (\ref{eq_nonlindyn_meth}) or (\ref{eq_nonlin_notripl}) by considering $\B x = (\theta_1, \theta_2, p_{\theta_1},p_{\theta_2})$ and rewriting the dynamical system as $\DB x = A \B x + C \B x^{\otimes 3}$, where

\begin{equation}
\B A = \begin{bmatrix}
0 & 0 & 2k & -3k \\
0 & 0 & -3k & 8k \\
3cg/l & 0 & 0 & 0 \\
0 & cg/l & 0 & 0
\end{bmatrix},
\end{equation}

and $\B C \in \RR^{4\times 64}$ with $C_{3, 41}=-6ck^2$, $C_{3, 42}=6ck^2$,$C_{3, 45}=25ck^2$,$C_{3, 46}=-25ck^2$,$C_{3, 61}=-24ck^2$,$C_{3, 62}=24ck^2$, $\B C_4 = -\B C_3$ and all the other entries set to zero, with $k = 6/(7ml^2)$ and $c = -1/2ml^2$.

\subsection{Connectivity density}\label{Methods:Connectivites}
Here we discuss the expected amount of connections based on the sparsity of the decoding matrix $\B D$ of a network implementing a generic dynamical system $\DB y = \B A \B y + \B B \B y \thinkron \B y$. 

\subsubsection{Fast connections \texorpdfstring{$\B \Omega_f$}{}}\label{Methods:FastCon}
For the fast connections, the connectivity matrix is given by $\B D\tran \B D$. 
For any pair of neurons $m,n$ we will have that a (fast) connection exists if $\B D_m\tran \B D_n \neq 0$, which means that if these two neurons ``share a dimension'' (i.e. $\B D_m,\B D_n$ have nonzero entries in at least one common spot and they are not orthogonal) they will need a fast connection among them. 
Let's denote with $0 \leq p_d^n \leq 1$ the probability that a neuron $n$ will participate in the representation of the $d-$th dimension (i.e. $p_d^n = P(\B D_n^d \neq 0)$). 
Let's assume that they are all independent. Then the probability that any given pair of neurons $n,m$ will need a connection is given by the probability that they both end up coding for at least one common dimension, given by
\[
p(\texttt{neurons } n,m \texttt{ code for common dimension}) = 1 - \prod_{d=1}^K (1 - p_d^n p_d^m).
\]
In the case where $p_d^n = p$ (such that neurons code for each possible dimension with equal probability) we can compute the expected number of fast connections for different neuron numbers and decoding densities as

\begin{equation}
\mathbb{E}(\# \texttt{fast connections}) = \frac{N(N-1)}{2} \left( 1 - (1-p^2)^K \right),
\end{equation}

%whose asymptotic behavior is depicted by the dashed line in Fig.~\ref{Fig3}A.

\subsubsection{Slow connections \texorpdfstring{$\B \Omega_s$}{}} \label{Methods:SlowCon}

Slow connections have the form $\B \Omega_s = \B D\tran(\B A + \lambda \B I)\B D = \B D\tran \B A \B D + \lambda \B D\tran \B D$. The second term, $\lambda \B D\tran \B D$, has exactly the form of the already considered case of fast connections.
We focus on the first term $\B D\tran \B A \B D$, which will add further connections to allow the network to solve linear dynamical systems. Since $\B A$ is not symmetric in general, $\B \Omega_s$ can be non symmetric too, hence the total possible number of slow connections is $N^2$, and will be so when the decoding matrix $\B D$ is not sparse.
If the matrix $\B A$ has a non-zero entry at a location $d,e$, all the neurons that code for dimension $d$ will have to connect to all the neurons that code for dimension $e$. The probability that two neurons $n,m$ will form a slow connection will be $p_d^n p_e^m$, or simply $p^2$ if the probability is uniform across dimensions and neurons. The expected number of slow connections (due to that single non-zero entry) is $ \sum_{n=1}^{N} \sum_{m=1}^N p_d^n p_e^m = (Np)^2 $,
where the last equality holds only in case of uniform probability. In that case we also have

\begin{equation}
\mathbb{E}(\# \texttt{slow connections}) \leq \mathbb{E}(\# \texttt{fast connections}) + N_A  (Np)^2,
\end{equation}
where $N_A$ is the number of nonzero entries in $\B A$. %The resulting relationship is plotted by the dashed line in Fig.~\ref{Fig3}B.

\subsubsection{Quadratic connections \texorpdfstring{$\B \Omega_{nl}$}{}}\label{Methods:QuadCon}

The quadratic connections take the form $\B \Omega_{nl} = \B D\tran \B B (\B D \thinkron \B D)$. If the decoding matrix is not sparse, the number of quadratic connections will be $\propto N^3$. In fact, the maximum possible number is given by $N^2(N-1)/2$, corresponding to each neuron ($N$) being connected to each possible pair ($\frac{N(N-1}{2}$). 
On the other hand, if $\B D$ is sufficiently sparse, we can reason as follows.
Denote with $G_d$ the group of neurons that code for dimension $d$, i.e. $G_d = \{n \ | \ \B D_n^d \neq 0\}$. Let’s assume that our dynamical system depends nonlinearly on dimensions $e$ and $f$, i.e. $\dot x_d \propto x_e x_f$, or equivalently $\B B_{d,eK+f} \neq 0$. Then, each neuron in $G_d$ needs to keep track of coincident firing of any neuron in $G_e$ with any other in $G_f$.  
The probability that a neuron in $G_d$ will need to take care of coincident spiking of the pair of neurons $m,n$ is $1 - (1- p_e^n p_f^m)(1 - p_e^m p_f^n)$, corresponding to the probability that at least one of the two neurons codes for dimension $e$ and the other for dimension $f$. In the case of uniform $p$ this reduces to $2p^2 - p^4$, so each neuron in $G_d$ will need an average of $\frac{N(N-1)}{2} (2p^2 - p^4)$ coincidence detectors, leading to an upper bound for the expected total number of multiplicative synapses 
\begin{equation}
\mathbb{E}(\# \texttt{multiplicative connections}) \leq N_B n_d \frac{N(N-1)}{2} (2p^2 - p^4) \approx N_B (N p)^3,
\end{equation}
where $n_d = \# G_d \approx Np$ and $N_B$ is the number of nonzero entries in $\B B$. The equality sign holds only in the case $N_B \leq 1$.  %The resulting relationship is plotted by the dashed line in Fig.~\ref{Fig3}C.

\subsubsection{Simulations}\label{Methods:Simulations}

In order to simulate the connectivity we fixed a decoder density $p$ and randomly filled the decoding matrix using a Bernoulli distribution $B(p)$ in each entry for 1000 times. For the fast connections we varied the size of the output signal - i.e. the size of the decoding matrix. For slow and multiplicative synapses the dimensionality of the signal $K$ did not affect the density of the resulting connections (not shown). What influenced the amount of slow and multiplicative synapses was the number of non-zero entries in the matrices $\B A$ and $\B B$, respectively.

\subsection{Code details}
Simulations were run in Ubuntu 20.04LTS on a Intel Core i5-6200U CPU with 32GB of RAM. 
The source code is available at \url{https://github.com/michnard/mult_synapses}.

\section*{Acknowledgements}
Version 4 of this preprint has been peer-reviewed and recommended by 
\emph{\PCI} (DOI link to the recommendation: \DOIrecommendationlink)
We thank Christian Machens and Nuno Calaim for useful discussions on the project.
This report came out of a collaboration started at the CAJAL Advanced Neuroscience Training Programme in Computational Neuroscience in Lisbon, Portugal, during the 2019 summer. The authors would like to thank the participants, TAs, lecturers, and organizers of the summer school. SWK was supported by the Simons Collaboration on the Global Brain (543009). WFP was supported by FCT (032077). MN was supported by European Union Horizon 2020 (665385). 

\section*{Conflict of interest disclosure}
The authors of this preprint declare that they have no financial conflict of interest with the content of this article.
%%%%%%%%%%%%% THE TEXT ENDS ABOVE THIS LINE %%%%%%%%%%%%%%%%%%%%%%%

\printbibliography[notcategory=ignore]

\section*{Supplementary Figures}
%%% reset figure counter
\setcounter{figure}{0}
%%% change name of figures to SUpp. Fig.
\renewcommand\thefigure{S\arabic{figure}}
\renewcommand{\theHfigure}{Supp.Fig.}
\renewcommand{\figurename}{Supp. Fig.}

\begin{figure}[h]
    \centering
    \includegraphics{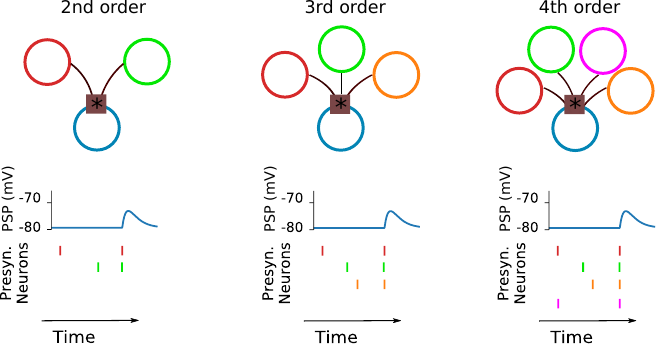}
    \caption{\textit{Higher order multiplicative interactions among cells.} Second (resp. third, fourth) order interactions require the post-synaptic cell to detect coincident activity in two (resp. three, four) pre-synaptic cells.}
    \label{SupFigHighOrder}
\end{figure}

\begin{figure}[h]
\centering
    \includegraphics[width=0.75\textwidth]{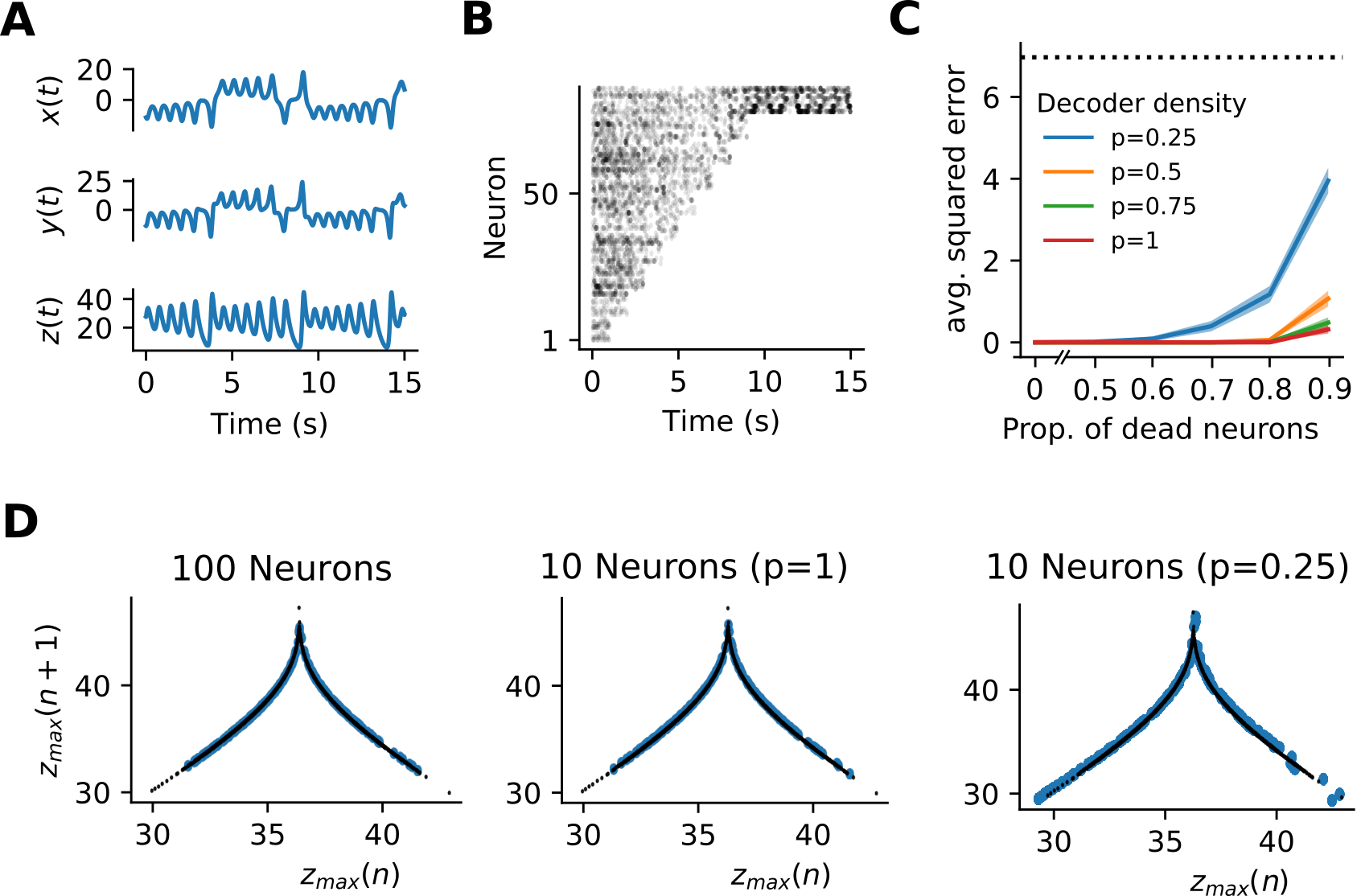}

\caption{\textit{Robustness of mSCNs.} 
\textbf{(A)} Example readout of a network of 100 neurons implementing a Lorenz dynamical system. For the first 9 seconds, 10 neurons were artificially killed every second. 
\textbf{(B)} Spike raster plot of the 100 neurons as a function of time. 
\textbf{(C)} Average squared error as a function of the proportion of neurons lost (out of 100 total) for different choices of decoder density. Shaded areas represent 95th confidence intervals. The error was computed by measuring the average squared distance of the network readout from the real solution of the Lorenz dynamical system (computed using a Runge-Kutta 4th order algorithm). The network was randomly initialized $1000$ times and the solution was approximated for $1$ second using $N=100, 90, 80, \dots, 10$ neurons, always starting from the same initial starting point. The dotted line represents the average squared error of an hypothetical constant readout center at the mean of the real solution in the $[0,1]$ time interval.
\textbf{(D)} Peak analysis on a 200 seconds network output using 100 cells (left), 10 cells and full decoder density (p=1, center), 10 cells and sparse decoder (p=0.25, right). Notice the loss of precision for the p=0.25 implementation.}
\label{SupFigRobust}
\end{figure}

\begin{figure}[h]
% \centering
\includegraphics[width=\textwidth]{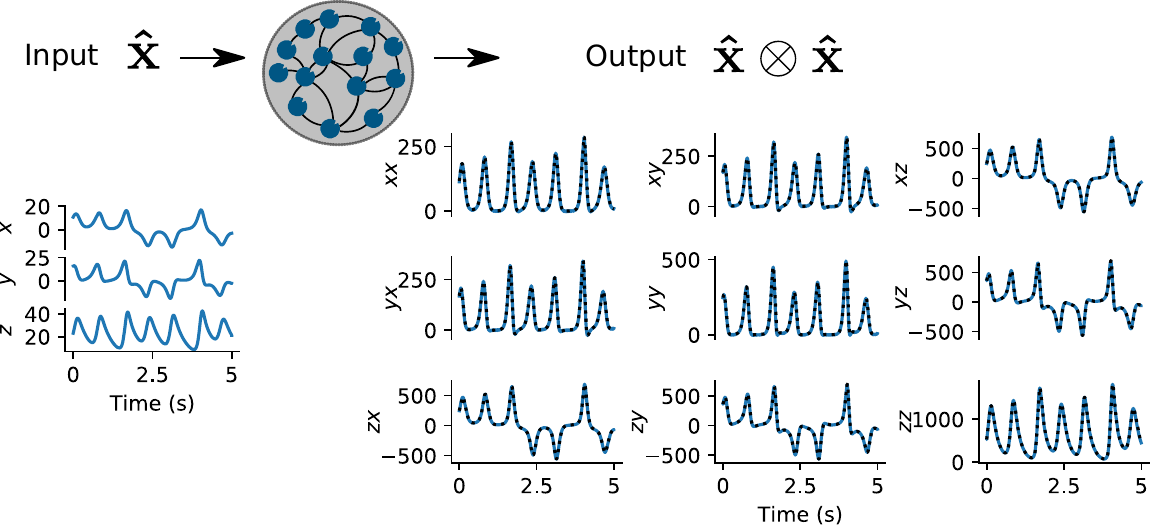}

    \caption{\textit{Representation of the Kronecker product of the input}. The Input $\HB x$ was given by a network which computed a Lorenz system. The second network, using eq. (\ref{eq_nonlindyn_meth}), outputs a signal $ \approx \HB x \thinkron \HB x$. Blue lines represent network output, black dotted lines represent the real $\HB x \thinkron \HB x$.}
\label{SupFigKronProd}
\end{figure}

\end{document}